\documentclass[preprint,12pt]{elsarticle}
\usepackage{amssymb}
\usepackage{amsthm}
\usepackage{graphicx}
\usepackage{amsfonts}
\usepackage{amsbsy}    
\usepackage{amsmath}
\usepackage{dcolumn}
\usepackage{bm}
\usepackage[applemac]{inputenc}
\usepackage{float}
\usepackage{hyperref}
\usepackage{graphics}
\journal{Physics Letters A}

\begin{document}
\begin{frontmatter}

\title{Fano effect and Andreev bound states in T-shape double quantum dots}

\author{A. M. Calle and M. Pacheco}
\address{Departamento de F\'{i}sica, Universidad T\'{e}cnica Federico Santa Mar\'{i}a - Casilla 110-V, Valpara\'{i}so, Chile}
  
\author{P. A. Orellana}  
\address{Departamento de F\'{\i }sica, Universidad Cat\'{o}lica del Norte - Casilla 1280, Antofagasta, Chile}
\begin{abstract}
In this paper, we investigate the transport through a T-shaped double quantum dot coupled to two normal metal leads left and right and a superconducting lead.  Analytical expressions of Andreev transmission and local Density of States  of the system at zero temperature have been obtained.  We study the role of the superconducting lead in the quantum interferometric features of the double quantum dot.   We report for first time the Fano effect produced by Andreev bound states in a side quantum dot. Our results show that as a consequence of quantum interference and proximity effect, the transmission from normal to normal lead exhibits Fano resonances due to Andreev bound states. We find that this interference effect allows us to study the Andreev bound states in the changes in the conductance between two normal leads.
\end{abstract}


\end{frontmatter}

Corresponding author:

e-mail : orellana@ucn.cl,

Telephone : +5655355507,

Fax : +5655355521

\section{Introduction}
Advances in techniques of nanostructures fabrication have made possible to built devices in which superconducting and normal metal nanostructures, such as quantum dots (QD), are connected \cite{hybrid-Kouwenhoven}. Interesting properties are opened when this combination of macroscopic phenomenon (superconductivity) and the ability to control single electrons in QD's system are considered. One of these properties is the so called proximity effect in which the superconducting-like properties may be induced in the normal metal \cite{Buzdin}. The most important characteristic of this effect  is the Andreev Reflection (AR), which has attracted much attention in normal metal-superconductor (N-S) junctions. In an AR process an electron with an energy below the superconducting gap is reflected at the interface as a hole while a Cooper pair of charge $2e$ is created in the superconducting side of the interface\cite{Andreev}. Andreev reflection plays an important role for the understanding of quantum transport properties of superconductor-normal metal systems. Moreover, in zero dimensional structures, as QDs, this process can give rise to discrete Andreev bound states (ABS). For instance, recently, Dirks et. al.\cite{Dirks} report transport measurements of ABS formed in a superconductor graphene quantum-dot normal system. They found  signature of ABS in the conductance peaks that occur inside the superconducting gap.

On the other hand, a wide variety of QDs systems such as T-shaped double quantum dot (DQD) systems, triple QD's, Aharonov Bohm QD rings, etc, have been studied recently~\cite{Tanaka,Trocha, Orellana,Tanamoto, Orellana06, Kobayashi04}. These systems exhibit interesting interference effects induced via multiple paths of electron propagation. One example is the Fano effect~\cite{fano}, which arises from the interference between a discrete state and the continuum, giving rise to characteristically asymmetric line-shapes. Unlike the conventional Fano effect in atomic physics, this effect in QD's system has the advantage in that its key parameters can be readily continuously tuned.
Theoretical studies in DQD systems  connected to two leads have been investigated considering both metallic leads~\cite{Tanaka, Zitko-T, Brown-T} and metal-superconducting leads~\cite{bai, Domanski-1}. For instance, recently, Doma\'nski $et$ $al$. studied a T-shaped double quantum dot coupled between metal and superconducting electrodes. These authors analyse the stability of Fano-type spectroscopic line-shapes with a decoherence induced by coupling to a floating lead. They conclude that decoherence has a detrimental effect on the quantum interferometric features~\cite{Domanski_Deco}. 

\begin{figure}[ht]
\begin{center}
\includegraphics[width=6cm]{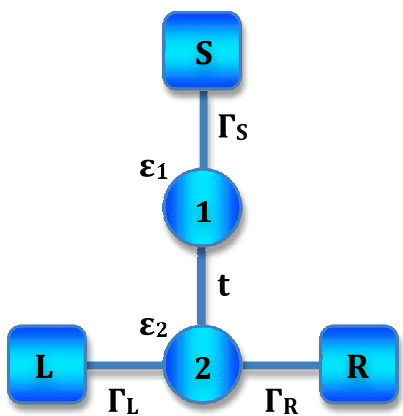}
\end{center}
\caption{Schematic view of T-shaped DQD system coupled to left (L) and right (R) normal leads and a superconductor lead (S) with an interdot coupling denoted by $t$.}
\label{fig1}
\end{figure}

In this work we report for  Fano-line shapes in the normal-normal transmission due to ABS in electronic transmission through a T-shaped DQD nanostructure coupled to two normal metal leads  and a third superconductor lead,  as sketched in fig.~ \ref{fig1}. This effect is studied as a function of the parameters defining the system.  We found that the Fano effect is robust against the DQD-superconductor lead coupling.  

This paper is organized as follows. In Sec. II we introduce the Hamiltonian of the system, and we adopt  the equation of motion approach for the Green functions in order to obtain expressions for the Andreev transmission, and for local density of states at zero temperature. In Sec. III we present the results for the transmission and density of states for different parameters. Finally in Sec. V we present our concluding remarks.

\section{Description of the Model}

We consider a T-shaped double QD system coupled with two normal leads (L and R) and one superconductor lead (S) as is shown in fig.~\ref{fig1}.  The full system is modelled by a non-interacting Anderson Hamiltonian, which can be written as:

\begin{equation}
\label{1}
H=H_{L(R)}+H_{S}+H_{dot}+H_{T},
\end{equation}
\noindent where $H_{L(R)}$ is the Hamiltonian for left and right normal leads given by
\begin{equation}
\label{2}
H_{\alpha}=\sum_{k,\sigma }\epsilon_{k \alpha}C^{\dag}_{\sigma k \alpha}C_{\sigma k\alpha}
\end{equation}
\noindent where $C^{\dag}_{\sigma k\alpha}$, $C_{\sigma k\alpha}$ are the creation and annihilation operator for electrons with momentum $k$ and spin $\sigma=\uparrow,\downarrow$ in the $\alpha=L(R)$ normal lead.  The standard BCS Hamiltonian for the superconductor lead is
\begin{equation}
\label{3}
H_{S}=\sum_{k,\sigma}\epsilon_{k S} C^{\dag}_{kS\sigma}C_{kS\sigma}+\sum_{k}\left(\Delta C^{\dag}_{kS\uparrow}C^{\dag}_{-kS\downarrow}+h.c.\right)
\end{equation}
\noindent where $\Delta$ is the energy gap for the superconductor which we assumed as a real parameter.  The Hamiltonian for the T-shaped DQD is
\begin{eqnarray}
\label{4 }
H_{dot}&=&\sum_{\sigma,i=1,2}\epsilon_{i}d^{\dag}_{i\sigma}d_{i\sigma}+\sum_{\sigma}t\left(d^{\dag}_{1\sigma}d_{2\sigma}+d^{\dag}_{2\sigma}d_{1\sigma}\right)
\end{eqnarray}
\noindent where $d^{\dag}_{i\sigma}$($d_{i\sigma}$) operator creates (annihilates) an electron in the $i$ quantum dot   with energy $\epsilon_{i}$ and inter-dot coupling $t$ which is taken as a real parameter.  Finally, the hybridization of the quantum dots with external reservoirs is given by
\begin{eqnarray}
\label{5}
H_{T}&=&\sum_{\sigma, k, \alpha}\left(V_{k \alpha}C^{\dag}_{\sigma k \alpha}d_{2\sigma}+h.c.\right) \nonumber \\&+&\sum_{k  \sigma}\left(V_{kS}C^{\dag}_{ kS \sigma}d_{1\sigma}+h.c.\right)
\end{eqnarray}
\noindent where, $V_{k\alpha}$ and $V_{kS}$ are the hopping between, left ($\alpha=L$) or right ($\alpha=R$)  lead and QD2, and between superconductor lead and  QD1 respectively.

The transport properties of the system are investigated by using the Green's functions formalism.  In order to obtain the Green's functions for the system, we use the method of Equation of Motion (EOM) where the Green's functions obtained by using this procedure can be written in a compact matrix form as
\begin{equation}
\label{6}
\boldsymbol{G}^{r}_{j,\sigma}=\boldsymbol{g}^{r}_{j,\sigma}+\boldsymbol{g}^{r}_{j,\sigma} \hspace{0.1cm} \boldsymbol{\Sigma}^{r}_{j} \hspace{0.1cm} \boldsymbol{G}^{r}_{j,\sigma}
\end{equation}
which is the Dyson equation with $g^{r}_{j,\sigma}$ the Green functions for a non-interacting QD and $\Sigma^{r}$ the retarded self-energy. 
At this point, it is useful to introduce the Nambu spinor notation in which the retarded and lesser Green's functions can be written  
\begin{equation}
\label{6.2}
G^{r}(t,t')=-\dot{\imath}\theta(t-t')\langle\Psi(t),\Psi^{\dag}(t')\rangle
\end{equation}
\begin{equation}
\label{9}
G^{<}\left(t,t'\right)=i\langle\Psi^{\dag}\left(t'\right),\Psi\left(t\right)\rangle
\end{equation}

\noindent with $\Psi^{\dag}_{1}=\left(d^{\dag}_{1\uparrow},d_{1\downarrow}\right)$ and $\Psi^{\dag}_{2}=\left(d^{\dag}_{2\uparrow},d_{2\downarrow}\right)$.
In the Nambu spinor space $g_{j}^{r}\left(\omega\right)$ can be written as

\begin{equation}
   \boldsymbol{\mathit{g}}_{1}^{r}\left(\omega\right) = \left(
      \begin{array}{cc}
        \frac{1}{\omega-\epsilon_{1}}  &  0    \\
0  &  \frac{1}{\omega+\epsilon_{1}}   \\
	\end{array}\right)  
\end{equation}

\begin{equation}
    \boldsymbol{\mathit{g}}_{2}^{r}\left(\omega\right) = \left(
      \begin{array}{cc}
        \frac{1}{\omega-\epsilon_{2}}  &  0    \\
0  &  \frac{1}{\omega+\epsilon_{2}}   \\
	\end{array}\right)  
\end{equation}

\noindent The retarded self energy is given by
\begin{equation}
 \boldsymbol{\mathit{\Sigma_{1}}}\left(\omega\right)=\sum_{\boldsymbol{\mathit{k},\alpha=L,R}}V_{\boldsymbol{\mathit{k\alpha}}} \hspace{0.1cm} \boldsymbol{\mathit{g}}_{\alpha}\left(\boldsymbol{\mathit{k}},\omega\right) \hspace{0.1cm} V_{\boldsymbol{\mathit{k\alpha}}}^{*}+t \hspace{0.1cm} \boldsymbol{\mathit{G}}_{2}\left(\omega\right) \hspace{0.1cm} t^{*}
\end{equation}

\begin{equation}
\boldsymbol{\mathit{\Sigma}}_{2}\left(\omega\right)=\sum_{\boldsymbol{\mathit{k}}}V_{\boldsymbol{\mathit{kS}}} \hspace{0.1cm} \boldsymbol{\mathit{g}}_{S}\left(\boldsymbol{\mathit{k}},\omega\right) \hspace{0.1cm} V_{\boldsymbol{\mathit{kS}}}^{*}+t \hspace{0.1cm}  \boldsymbol{\mathit{G}}_{1}\left(\omega\right) \hspace{0.1cm} t^{*}
\end{equation}

Under the wide-bandwidth approximation we have

\begin{equation}
\sum_{\boldsymbol{\mathit{k}}}V_{\boldsymbol{\mathit{k\alpha}}} \hspace{0.1cm} \boldsymbol{\mathit{g}}_{\alpha}\left(\boldsymbol{\mathit{k}},\omega\right) \hspace{0.1cm} V_{\boldsymbol{\mathit{k\alpha}}}^{*}=-\dot{\imath}\hspace{0.1cm} \frac{\Gamma_{\alpha}}{2} \left(
      \begin{array}{cc}
       1  &  0    \\
0  &  1   \\
	\end{array}\right)  
\end{equation}

\begin{equation}
\sum_{\boldsymbol{\mathit{k}}}V_{\boldsymbol{\mathit{kS}}} \hspace{0.1cm} \boldsymbol{\mathit{g}}_{S}\left(\boldsymbol{\mathit{k}},\omega\right) \hspace{0.1cm} V_{\boldsymbol{\mathit{kS}}}^{*}=\dot{\imath}\hspace{0.1cm} \frac{\Gamma_{S}}{2}g\left(\omega\right)\left(
      \begin{array}{cc}
       1  &  -\frac{\Delta}{\omega}    \\
-\frac{\Delta}{\omega}  &  1   \\
	\end{array}\right)  
\end{equation}

Where, we use the notation 
\begin{equation} 
g\left(\omega\right) = -\left[ \frac{\theta\left(\Delta-|\omega|\right)}{\sqrt{\Delta^{2}-\omega^{2}}}+ \dot{\imath}\hspace{0.1cm} \textrm{sgn}\left(\omega\right) \frac{\theta\left(|\omega|-\Delta\right)}{\sqrt{\omega^{2-}\Delta^{2}}} \right] 
\end{equation}

\noindent Then, retarded Green's functions can be determined from the following set of coupled equations
\begin{eqnarray}
\boldsymbol{G}_{1}\left(\omega\right)^{-1}&=&\left[\omega+\dot{\imath}\hspace{0.1cm} \frac{g\left(\omega\right)\Gamma_{S}}{2}\right]\textbf{I}-\epsilon_{1}\boldsymbol{\sigma}_{z}\nonumber\\ &+& \dot{\imath}\hspace{0.1cm} \frac{g\left(\omega\right)\Gamma_{S}\Delta}{2\omega}\boldsymbol{\sigma}_{x}-t^{2}\boldsymbol{G}_{2}\left(\omega\right) ,
\end{eqnarray}
\begin{equation}
\boldsymbol{G}_{2}\left(\omega\right)^{-1}=\left[\omega+\dot{\imath}\hspace{0.1cm} \frac{\Gamma_{L}+\Gamma_{R}}{2}\right]\textbf{I}-\epsilon_{2}\boldsymbol{\sigma}_{z}-t^{2}\boldsymbol{G}_{1}\left(\omega\right)
\end{equation}
\noindent where \textbf{I} is the identity matrix and $\boldsymbol{\sigma}_{x}$, $\boldsymbol{\sigma}_{z}$ denote the usual Pauli matrices.

The electronic current in the lead $L$ can be calculated with the equation:
\begin{equation}
\label{7}
I_{L}=-e\hspace{0.1cm}\langle \frac{dN_{L}}{dt}\rangle
\end{equation}
with $N_{L}=\sum_{k\sigma}C^{\dag}_{kL\sigma}C_{kL\sigma}$.  The current is then \cite{Kim-etal}
\begin{equation}
\label{8}
I_{L}=2e\sum_{k}\textrm{Im} \hspace{0.1cm} \textrm{Tr} \hspace{0.1cm}V^{\dag}_{kL}G^{<}_{\sigma,k\sigma}\left(t,t\right)
\end{equation}
where $G^{<}_{\sigma,k\sigma}(t,t)$ is the lesser Green's function. 

So, in order to obtain the transmission probability from left to superconducting lead we use eq.~(\ref{8}) and we get after some algebraic manipulation
\begin{equation}
\label{10}
I_{A}=\frac{2e}{h}\int d\omega \hspace{0.1cm} T_{A}\left(\omega\right)\left[f\left(\omega\right)-f\left(\omega-\mu_{L}\right)\right] ,
\end{equation}
where $T_{A}\left(\omega\right)$ is the Andreev transmission probability from the  left to superconducting lead. 

The Andreev transmission, in which an electron that comes from the left will be reflected as a hole creating an extra Cooper pair in the superconducting lead, is given by
\begin{equation}
T_{A}\left(\omega\right)=\Gamma^{2}_{L}\left[\hspace{0.1cm} \arrowvert G_{2,12}\left(\omega\right)\arrowvert^{2}+\arrowvert G_{2,12}\left(-\omega\right)\arrowvert^{2}\right] ,
\end{equation}
where the index $j$ in the Green function $G^{r}_{j,\alpha\beta}$ denotes the QD site while the indices $\alpha$, $\beta$ denote the elements of the $2\times2$ matrix in the Nambu spinor space.  If we consider the approximation $\omega \ll \Delta$ where only the off-diagonal terms of the matrix $\Sigma_{S}$ are different from zero and have a constant value, the Andreev Transmission can be written in simplified form as

\begin{equation}
T_{A}= \frac{\Gamma_{L}}{D\left(\omega\right)} \frac{t^{4}\left(\omega+\epsilon_{1}\right)^{2}\left(\omega+\epsilon_{2}\right)^{2}}{\left(\omega-\epsilon_{1}\right)^{2}\left(\omega-\epsilon_{2}\right)^{2}}\left(\frac{\Gamma_{s}}{2}\right)^{2} ,
\end{equation}

where, $D\left(\omega\right)$ is given by

\begin{eqnarray}
D\left(\omega\right)&=&\Bigg[\left( \omega^{2}-\epsilon_{1}^{2}-\left(\frac{\Gamma_{s}}{2}\right)^{2} \right)\left( \omega^{2}-\epsilon_{2}^{2}-\left(\frac{\Gamma_{d}}{2}\right)^{2} \right)\nonumber\\&+&2t^{2}\left(\omega^{2}+\epsilon_{1}\epsilon_{2}\right)+t^{4}\Bigg]^{2} \nonumber\\
&+& \hspace{0.2cm} \Gamma_{d}^{2}\omega^{2}\left[ \omega^{2}-\epsilon_{1}^{2}-\left(\frac{\Gamma_{s}}{2}\right)^{2}-t^{2}\right]^{2} .
\end{eqnarray}

On the other hand, to obtain the transmission probability from $L$ to $R$ normal leads for spin up we use the Landauer formula \cite{Landauer}, from which we obtain:
\begin{equation}
T_{LR}\left(\omega\right)=\Gamma_{L}\Gamma_{R}\left[\arrowvert G_{2,11}\left(\omega\right)\arrowvert^{2}+\arrowvert G_{2,12}\left(\omega\right)\arrowvert^{2} \right]
\end{equation}

which yields to the following equation
\begin{eqnarray}\nonumber\\
T_{LR}\left(\omega\right)&=&\frac{\Gamma_{L}\Gamma_{R}}{D\left(\omega\right)}\left(\left[ \omega^{2}-\epsilon_{1}^{2}-\left(\frac{\Gamma_{s}}{2}\right)^{2} \right]^{2}\left[ \left(\omega+\epsilon_{2}\right)^{2}+\left(\frac{\Gamma_{d}}{2}\right)^{2}  \right]\right. \nonumber\\&-& \left. 2t^{2}\left(\omega-\epsilon_{1}\right)\left(\omega+\epsilon_{2}\right)\left[\omega^{2}-\epsilon_{1}^{2}-\left(\frac{\Gamma_{s}}{2}\right)^{2}\right] \right. \nonumber\\ &+& \left. t^{4}\left(\omega-\epsilon_{1}\right)^{2} + t^{4}\frac{\left(\omega+\epsilon_{1}\right)^{2}\left(\omega+\epsilon_{2}\right)^{2}}{\left(\omega-\epsilon_{1}\right)^{2}\left(\omega-\epsilon_{2}\right)^{2}}\left(\frac{\Gamma_{s}}{2}\right)^{2} \right)
\end{eqnarray}

\section{Results} 

\begin{figure}[t]
\begin{center}
\includegraphics[width=10cm]{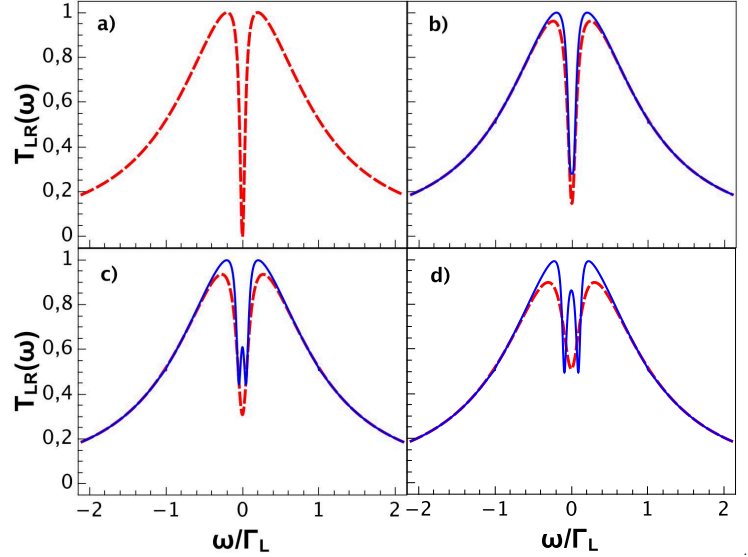}
\end{center}
\caption{\label{fig2}Transmission from left to right lead versus energy for a T-shaped DQD for a) $\Gamma_S=0$ b) $\Gamma_S=0.05\Gamma_L$, c) $\Gamma_S=0.1\Gamma_L$ and d)$\Gamma_S=0.2\Gamma_L$.  Solid line corresponds to $\Delta=1\Gamma_{L}$ and dashed line to $\Delta=0$}
\end{figure}

In what follows the energies will be given in  units of the parameter $\Gamma_{L}$ and we set the site energies as $\epsilon_{1}=\epsilon_{2}=0$.
In order to analyze the effect of the superconducting lead in the quantum interferometric features of a T-shaped DQD,  fig.~ (\ref{fig2})  displays the transmission from L to R for $\Delta=\Gamma_{L}$ and weak inter-dot coupling $t=0.2$ $\Gamma_{L}$ for different values of the dot-superconductor lead coupling $\Gamma_S$.  fig.~ (\ref{fig2}) a)  corresponds to the case without superconductor lead ($\Gamma_S=0$).  In this case the transmission from L to R shows a symmetric Fano line-shape and vanishes at $\epsilon=0$. This effect appears due to the quantum interference between a localized state in the side quantum dot with the continuum of the central quantum dot coupled to the leads. In the  presence of the third superconducting lead ($\Gamma_S\neq 0$), the zero in the transmission becomes in two small antiresonances as is shown in fig.~(\ref{fig2}) c) and d). For comparison, we show the effect of a third normal lead ($\Delta=0$, red dashed line). In this case for sufficiently large  $\Gamma_S$, the initial antiresonance in the transmission disappears. This implies that the third normal lead introduces decoherence effects destroying the Fano line-shape, as it is expected for a DQD system coupled to three normal contacts  \cite{Wen-Zhu}.  Therefore, as $\Gamma_S$ increases, the presence of the superconductor lead induces  that the initial Fano-antiresonance in the transmission splits in two antiresonances.

At this point is important to mention that a similar characteristic has also been studied in hybrid systems for  Doma\'nski $et$ $al$. \cite{Domanski_Deco}, by using a system formed by N-DQD-S. They showed  that a third normal  lead introduces decoherence that suppresses the Fano effect concluding that decoherence has a detrimental effect on quantum interferometric features. The Fano lineshapes seem to be rather fragile entities with respect to the third normal lead.  
\\

\begin{figure}[ht]
\begin{center}
\includegraphics[width=10cm]{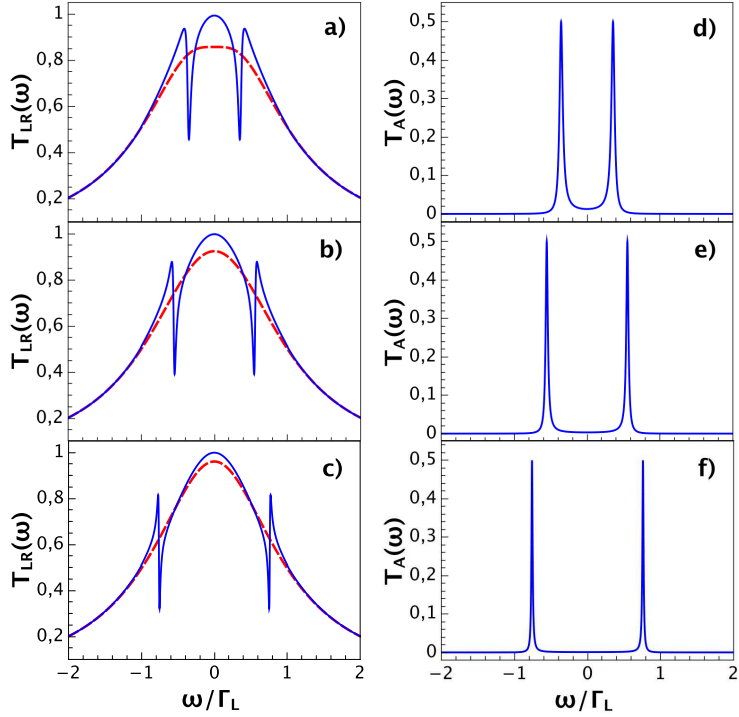}
\end{center}
\caption{\label{fig3}Transmission from left to right lead (left panel) and Andreev transmission (right panel) versus energy for a T-shaped DQD with parameters: $\Delta=1\Gamma_{L}$, $\Gamma_{R}=\Gamma_{L}$, $t=0.2\Gamma_{L}$, $\epsilon_{1}=\epsilon_{2}=0$ and $\Gamma_S=1\Gamma_{L}$ for a) and d), $\Gamma_S=2\Gamma_L$ for b) and e)  and $\Gamma_S=4\Gamma_L$ for c) and f). Dashed line in left panel corresponds to transmission from L to R with $\Delta=0$, i.e for a system with three normal leads.}
\end{figure}

In order to get a better understanding of the above behaviour, in fig.~(\ref{fig3}) a), b) and c) we plot the transmission from L to R versus energy  as $\Gamma_{S}$ increases.  Additionally, in the right panel in fig.~(\ref{fig3}) d), e) and f) we show the Andreev transmission. We can observe how the transmission evolves while $\Gamma_{S}$ increases.  The transmission from L to R  shows two well defined Fano antiresonances with an imaginary asymmetry parameter.  On the other hand, the Andreev contribution is appreciable only inside the superconducting gap.  Note that Andreev transmission is symmetric $T_{A}\left(\omega\right)=T_{A}\left(-\omega\right)$ due to it involves both the particle and hole degrees of freedom.

Therefore, it is interesting to contrast Domanski's results for a T-shape DQD with a normal floating third lead with our results for a system with a superconducting third lead. We note that as the coupling with the superconducting lead is augmented, the initial drop in the Fermi energy in the transmission from L to R, begin to split in two small antiresonances. As $\Gamma_{S}$ is increased  these antiresonances become Fano line-shapes,   as we can see in Fig. (\ref{fig3}) a), b) and c).  The Fano antiresonances are localized at $E\approx\pm\left(a\sqrt{\Gamma_{S}}+b\Gamma_{S}+c\right)^{2}$ with $a=0.612$, $b=-0.109$, $c=0.086$.  Their position in energy  depend on dot-superconductor coupling $\Gamma_{S}$ due to the proximity effect.  They have identical shapes but an opposite sign of the imaginary asymmetry parameter $q$.  It is important to note in fig.~(\ref{fig3}) (d), e) and f))  that the resonances exhibited in the Andreev transmission are centered in the same position as the Fano antiresonances in the transmission from L to R.  Therefore, this suggests, that the Fano antiresonances in the normal transmission are due to the ABS.

\begin{figure}[t]
\begin{center}
\includegraphics[width=8cm]{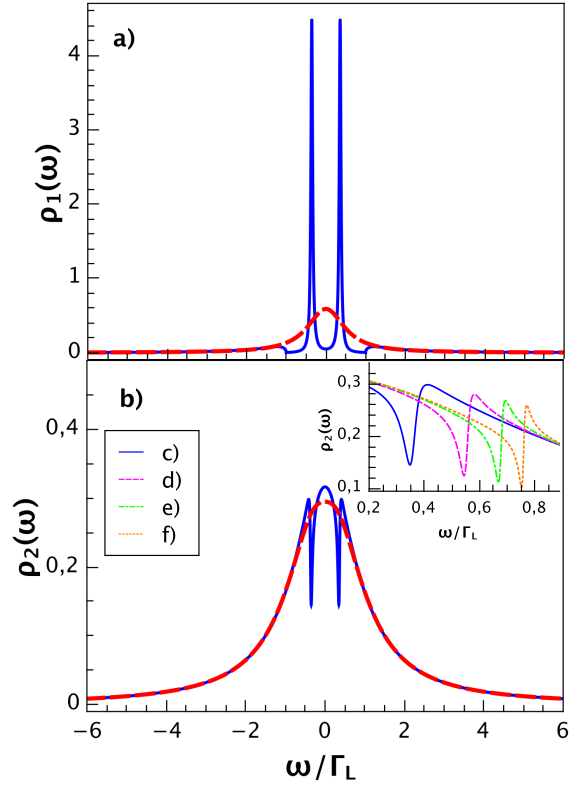}
\end{center}
\caption{\label{fig4}Local density of states for a) QD 1 and b) QD 2, where the solid line corresponds to $\Delta=1\Gamma_{L}$ while dashed line corresponds to $\Delta=0$.  With parameters $\Gamma_{R}=\Gamma_{L}$ and $\Gamma_{S}=\Gamma_{L}$.  The inset in b) shows the evolution of Fano antiresonances with $\Gamma_{S}$ for c)$\Gamma_S=1\Gamma_L$ d) $\Gamma_S=2\Gamma_L$,  e) $\Gamma_S=3\Gamma_L$, and f)$\Gamma_S=4\Gamma_L$.}
\end{figure}

Finally, we illustrate the density of states $\rho_{1}\left(\omega\right)$ and $\rho_{2}\left(\omega\right)$ for quantum dots 1 and 2 respectively.  In fig.~(\ref{fig4}) a) we can see how the Lorentzian curve centered in $\epsilon_{2}$ (dashed line) splits in two peaks (solid line), due to the ABS.  On the other hand, in fig.~(\ref{fig4}) b) we observe the appearing of two antiresonances, as $\Gamma_{S}$ increases.  In the inset in fig.~(\ref{fig4}) b) we shows how the superconducting lead does not introduce decoherence and the Fano effect remains as the dot-superconductor lead coupling $\Gamma_{S}$ is augmented, as we mentioned above.

We expect that the Fano-Andreev effect remains valid even if the electron-electron interaction is taken into account. In fact, in embedded coupled QD, the main effect of the electron-electron interaction is to shift and to split the resonance positions \cite{Sztenkiel, Aono, Lu}.  This occurs because the on-site Coulomb repulsion energy $U$ introduces a renormalization of the site energies.  In analogy with coupled QD, we expect that, depending on the relation between the interdot coupling and the on-site Coulomb interaction, different regimes arise.	For $t /U\gg1$, 	the resonances and antiresonances would split into two,  separated by the on-site Coulomb energy, while for $t/ U\ll1$, the resonances and antiresonances would	occur in pairs. Moreover, it was demonstrated that the Fano effect is robust against e-e interaction even in Kondo regime  \cite{Johnson, Sato}. Therefore we expect that electron-electron interaction does not break the Fano-Andreev effect. 

\section{Conclusions}
In the present work, we have studied the transport properties of a DQD coupled in T-shape configuration to the conducting and superconducting leads.  We studied transport properties of the system, such as the normal-normal transmission, Andreev transmission and local density of states.  The role of the superconducting lead in the quantum interferometric features is also analysed. The normal-normal leads transmission displays two Fano line-shapes characterized by an opposite sign of the imaginary asymmetry parameter $q$.  We show that Fano antiresonances in the normal transmission are due to the Andreev reflections in the superconducting lead. It is important to mention that this effect survives even for strong dot-superconductor lead coupling. This result open new possibilities to study ABS in electronic transport through quantum dots.

\section{Acknowledgments}
We acknowledge the financial support from FONDECYT program Grants No. 1100560 and No. 1100672, and PIIC-UTFSM 2012.

\end{document}